\UseRawInputEncoding
\documentclass[aps,prl,twocolumn,showkeys,longbibliography,superscriptaddress,footinbib]{revtex4-2}

\usepackage{mathtools} 

\usepackage[normalem]{ulem} 
\usepackage{amsfonts}
\usepackage{comment}
\usepackage{amsmath}
\usepackage{physics}
\usepackage{graphicx}
\usepackage{subfigure}
\usepackage{makecell,multirow}
\usepackage{babel}
\usepackage{amstext}
\usepackage{esint}
\usepackage[unicode=true,pdfusetitle,
bookmarks=true,bookmarksnumbered=false,bookmarksopen=false, breaklinks=false,pdfborder={0 0 1},backref=false,colorlinks=false]
 {hyperref}
\usepackage{pdfpages}	
\usepackage{pgffor}		
\usepackage{upgreek}
\usepackage{braket}
\usepackage{cleveref}
\usepackage{cancel}

\newcommand{\blue}[1]{{\color{blue} #1}}

\makeatletter
\AtBeginDocument{\let\LS@rot\@undefined} 
\makeatother							 

\renewcommand{\section}[1]{{\it {#1}. }} 

\crefname{equation}{Eq.}{Eqs.}
\crefname{figure}{Fig.}{Figs.}

\begin{document}

\title{Spin-chain multichannel Kondo model via image impurity boundary condition}

\author{Jordan Gaines}
\affiliation{%
 Department of Physics and Astronomy, Purdue University, West Lafayette, Indiana 47907, USA.
 }

\author{Guangjie Li}
\email{guangjie.li@utah.edu}
\affiliation{%
Department of Physics and Astronomy, Purdue University, West Lafayette, Indiana 47907, USA.
}
\affiliation{%
Department of Physics and Astronomy, University of Utah, Salt Lake City, Utah 84112, USA.
}

\author{Jukka I. V{\"a}yrynen}
\email{vayrynen@purdue.edu}
\affiliation{%
Department of Physics and Astronomy, Purdue University, West Lafayette, Indiana 47907, USA.
}

\date{\today}

\begin{abstract}
One of the signature observables for the electronic multichannel Kondo model is the impurity entropy, which was found in $J_1$-$J_2$ Heisenberg chains with the open boundary condition (OBC) and periodic boundary condition (PBC), for the one-channel and two-channel cases respectively.
However, it is not clear how to generalize OBC and PBC in Heisenberg chains to find the multichannel Kondo impurity entropy with more than two channels.
In this paper, we demonstrate that the correct boundary condition for realizing multichannel Kondo physics in Heisenberg chains is the image impurity boundary condition (IIBC) which preserves reflection symmetry and yields the expected impurity entropy, $\ln[(\sqrt{5}+1)/2]$ for the three-channel case and $\ln\sqrt{3}$ for the four-channel case. Moreover, the IIBC reduces to OBC for the one-channel case and to PBC for the two-channel case. With IIBC, the finite-size scaling of the impurity entropy and the total impurity spin match the finite-temperature corrections in the electronic multichannel Kondo model. Additionally, we show  dependence of the impurity entropy, the total impurity spin, and their scaling behaviors on the XXZ anisotropy $\Delta$ (equivalently the Luttinger liquid parameter), revealing impurity physics in a multichannel Luttinger liquid.
\end{abstract}

\maketitle

\section{\blue{Introduction}}~The electronic Kondo effect occurs at low temperatures when conduction electrons screen a magnetic impurity leading to a many-body entangled state \cite{kondo1964,hewson1997heavyfermionkondo}.  This model was later generalized to consider multiple free-fermion channels overscreening the impurity \cite{nozieres1980mck}, and this generalization has proved to be a useful model to illustrate exotic many-body phenomena~\cite{cox1998exotickondometals,ludwig1994field,affleck2010quantum}, especially in mesoscopic devices \cite{potok2007observation2ck, keller2015universalfermiliquid, iftikhar20152ckchargestates, iftikhar2018ballisticchargekondo, pouse2023kondocircuit}, many of which exhibit non-Fermi liquid behavior. 
A striking example is the zero-temperature impurity entropy $S_{\text{imp}} = \ln g$ where $g$ plays the role of a  ``ground-state degeneracy'' which is generally not an integer~\cite{affleck1992noninteger,ludwig1994field,
affleck2010quantum}.  For a Kondo model with $M$ channels, it has been shown through the Bethe Ansatz method~\cite{destri1984mck, tsvelick1985thermokondo} and by boundary conformal field theory methods \cite{affleck1992noninteger,ludwig1994field,affleck1995cftkondo,affleck2010quantum,kimura2021kondoabcd} that: 
\begin{equation}
\label{eq:g_MCK} 
    g(M) = 2\cos\left(\frac{\pi}{M+2}\right) .
\end{equation} 
Note that $g(M=1)=1$ gives zero entropy due to exact screening and $g(M=2)=\sqrt{2}$ (the quantum dimension of a free Majorana) gives $\ln(2)/2$ entropy due to overscreening~\cite{emery19922ckresonantlevel}. 

Recently, the multichannel Kondo model has also attracted the attention of the topological quantum computation community since $g(M)$ matches the quantum dimension of various non-Abelian anyons~\cite{lopes20203ckanyons,komijani2020kondoanyons}, offering the tempting interpretation that an emergent anyon appears at the overscreened impurity. 
The case of $M=3$ is of particular interest since $g(M=3) = \varphi \approx1.618$, the golden ratio, hints at the existence of Fibonacci anyons \cite{lopes20203ckanyons, komijani2020kondoanyons, kimura2021kondoabcd, han2022kondoentropy, RenTQCchiralKondochain2024}, exotic quasiparticles that may provide a path to universal
topological quantum computation~\cite{nayak2008topoanyons}. However, the precise connection to anyons remains unclear and, so far, an explicit construction is only known for the Majorana case $M=2$~\cite{emery19922ckresonantlevel,sengupta19942ckanyons}, motivating further investigation of $g(M>2)$ in multichannel Kondo models. Beyond the pure multichannel Kondo models, Ref.~\cite{KarkiZ3parafermion2023} explicitly finds the $\mathbb{Z}_3$ parafermion in the two-impurity Kondo model~\cite{pouse2023kondocircuit}.

While experiments towards observing impurity entropy and verifying Eq.~(\ref{eq:g_MCK}) are progressing \cite{hartman2018directentropymeasurement, child2022qdotentropymeasurement, han2022fractionalentropy}, this formula has been numerically verified for up to $M=4$ channels via various numerical renormalization group (NRG) methods \cite{WEICHSELBAUM20122972,MitchellNRG2014,matan2022electronic3ck,matan2023chiralnrg}. 
Theoretically, conformal field theory methods, specifically non-abelian bosonization, have shown that the free fermions of the electronic $M$-channel Kondo model can be bosonized using the ``charge" [$U(1)_{2M}$], ``spin" [$\text{SU}(2)_{M}$] and ``channel" [$\text{SU}(M)_2$] currents in the corresponding Kac-Moody algebras~\cite{affleck1995cftkondo,ludwig1994field,kimura2021kondoabcd}. The Kondo interaction is only coupled with the spin sector, suggesting that the fermionic background is not necessary and that Kondo physics can also be studied in systems such as spin chains~\cite{laflorencie2008kondo,SU2kspinchain,Zhakenovimpurity} or quantum circuits~\cite{sarma2025designbenchmarksemulatingkondo}.

Furthermore, it has been widely known that the low-energy effective theory of the spin-chain Heisenberg (XXX) model is described by the $\text{SU}(2)_1$ Kac-Moody algebra~\cite{GarateAffleckPhysRevB.81.144419}. Indeed, the one-channel and two-channel Kondo impurity entropies, that is Eq.~(\ref{eq:g_MCK}) at $M=1,2$, were observed in the open boundary and periodic boundary XXX spin-chain models, respectively \cite{laflorencie2008kondo,alkurtass2016spin2CK}. 
These spin-chain models are of interest because they directly demonstrate that a fermionic background is not necessary to realize Kondo physics. Additionally, the use of a spin-chain model enables us to take advantage of the DMRG algorithm \cite{white1992dmrg}, which has seen great success in quantum simulation \cite{schollwoeck2011dmrgreview, baiardi2020chemdmrg,wu2025}. Spin-chain models also complement fermionic models simulated by NRG. However, as we illustrate below, it is not obvious how to extend the spin chain formulation to $M > 2$ channels. 
This extension is one of the key results in this letter and serves to suggest that the physics of the Kondo effects may be bosonic.

Besides the $\text{SU}(2)$-symmetric case, we also consider the effects of exchange anisotropy on the spin-chain multichannel Kondo models.
One-channel \cite{kattel2024xx1ck} and two-channel Kondo physics \cite{tang2025xx2ck} have also been observed in XX Heisenberg chains where the next-nearest neighbor coupling $J_2$ is not needed, unlike the isotropic case \cite{laflorencie2008kondo, alkurtass2016spin2CK}.
Notably, the absence of the next-nearest neighbor interaction makes numerical simulation much easier, motivating potential investigation into the viability of XX chains for simulating Kondo physics \cite{CRAMPE2013526}. 
Furthermore, one-channel Kondo physics has also been demonstrated for a general anisotropic 1D antiferromagnetic Heisenberg chain \cite{furusaki1998kondoXXZchain}. Following these investigations, we introduce anisotropy into our multichannel Kondo model, parametrized by $\Delta$ with $\Delta = 1$ being the isotropic model and $\Delta = 0$ being the XX model. 
We numerically observe that $g\sim d^{(M-1)/2}$ where $d = 1-\arccos(\Delta)/\pi$ (see Fig.~\ref{fig3}). Here, anisotropy is related to  the effective Luttinger liquid parameter of the XXZ chain $K=1/(2d)$~\cite{giamarchi2003quantum}, motivating further investigations on the entanglement in multichannel Luttinger liquid impurity models~\cite{KaneFisher1992,GiulianoKondoweaklinkSpinChain2018,
TanEntanglementflowKaneFisher2025,KattelXXZ2025,RylandsQIluttinger2016,Sirker2008,Collura2013,ZhaoCriticalXXZdefects2006}.

\section{\blue{The Multichannel Spin-Chain Kondo Model}}~We start by introducing our model to realize multichannel Kondo physics using spin chains.  We consider a spin-$1/2$ chain geometry composed of $M$ chains of length $L$, one chain for each channel, coupled at the left by an impurity spin $\vb*{\mathrm{S}}_0$ and at the right by an image impurity spin $\vb*{\mathrm{S}}'_0$. 
The spins are coupled via the next-nearest neighbor antiferromagnetic Heisenberg interaction.  We will start by considering the isotropic case (XXX model) and then introduce anisotropy later (XXZ model).  Concretely, we consider the following Hamiltonian:
\begin{align}
\label{eq:MCK_ham}
&H =\ H_{\text{imp}} + H_{\text{image}} + \sum_{k=1}^M H^k_{\text{bulk}}\,, \\
\label{eq:ham_bulk}
&H^k_{\text{bulk}} \! = \! \  J_1 \sum_{j=1}^{L-1} \vb*{\mathrm{S}}^{k}_{j} \cdot \vb*{\mathrm{S}}^{k}_{j+1} + J_2 \sum_{j=1}^{L-2} \vb*{\mathrm{S}}^{k}_{j} \cdot \vb*{\mathrm{S}}^{k}_{j+2} \,,   \\
\label{eq:ham_imp}
&H_{\text{imp}}\! = \! \ J_K \! \sum_{k=1}^M \Big[\vb*{\mathrm{S}}_0 \! \cdot \! \qty(\vb*{\mathrm{S}}^{k}_{1} \!+\! \frac{J_2}{J_1} \vb*{\mathrm{S}}^{k}_{2}) \!+\! \frac{J_2}{J_1} \sum_{k'>k} \vb*{\mathrm{S}}^{k}_{1} \! \cdot \! \vb*{\mathrm{S}}^{k'}_{1} \Big],\\
\label{eq:ham_image}
&H_{\text{image}}\! = \! \ J_K' \! \sum_{k=1}^M \Big[\vb*{\mathrm{S}}_0' \! \cdot \! \qty(\vb*{\mathrm{S}}^{k}_{L} \!+\! \frac{J_2}{J_1} \vb*{\mathrm{S}}^{k}_{L-1}) \!+\! \frac{J_2}{J_1} \sum_{k'>k} \vb*{\mathrm{S}}^{k}_{L} \! \cdot \! \vb*{\mathrm{S}}^{k'}_{L} \Big],
\end{align}
where $\vb*{\mathrm{S}}^{k}_{j}$  is the spin-1/2 operator  $\vb*{\mathrm{S}}=\hbar (\sigma^x, \sigma^y, \sigma^z)/2 $ for the $j^{\text{th}}$ spin in $k^{\text{th}}$ channel. $H_{\text{imp}}$ and $H_{\text{image}}$ implement interaction with and across the impurity and image impurity respectively. The nearest-neighbor couplings between spins are also illustrated in \cref{fig1}(a). For simplicity, we work in units $J_1 = \hbar = 1$. 
To implement anisotropy, we use the Hamiltonian \cref{eq:MCK_ham} but replace $ \vb*{\mathrm{S}}_i \cdot \vb*{\mathrm{S}}_j \to \mathrm{S}_i^x \mathrm{S}_j^x + \mathrm{S}_i^y \mathrm{S}_j^y + \Delta \mathrm{S}_i^z \mathrm{S}_j^z $. We consider only the critical (gapless) regime $-1\!<\!\Delta\!\le\!1$~\cite{giamarchi2003quantum}. The isotropic case $\Delta\!=\!1$ gives the multichannel spin-chain Kondo model where the Kondo impurity entropy matches Eq.~(\ref{eq:g_MCK}). Note that the $\Delta=0$ case is a free-fermion model only for $M=1,2$, which further demonstrates that the generalization of \cref{fig1}(a) is nontrivial.

The next-nearest neighbor couplings $J_2$ are introduced to suppress a marginally irrelevant bulk interaction which contributes to finite size effects \cite{nomura1993phase,nomura1994critical,eggert1996numericalj2,laflorencie2006boundarycritical,alkurtass2016spin2CK}. 
These finite size effects are minimized when the marginal coupling constant in the bulk vanishes, which has been shown to occur at the critical point $J^c_2 \approx 0.2412$ \cite{eggert1996numericalj2} between the gapless ($J_2 < J^c_2$) and dimerized ($J_2 > J^c_2$) phases.
Note further that we fix the chain length $L$ to be even as it has been shown that an odd chain length leads to quantitatively different entanglement entropy profiles which do not match Kondo physics \cite{sorensen2007impurityentanglement, affleck2009impurityentropy, sorensen2007kondoimpurity}.

\section{\blue{The Subtle $M=2$ Case}}~Even the realization of the two-channel model~[see \cref{fig1}(a)] using an $\text{SU}(2)_1$ Heisenberg chain is subtle. 
Following Ref.~\cite{alkurtass2016spin2CK}, we consider a uniform, periodic chain of length $2L+2$ ($J_K=J'_K=J_1$) to extract the impurity entropy at the Kondo intermediate coupling fixed point.
\begin{figure}[tb]
\centering
\includegraphics[width=0.95\columnwidth]{PRLv2spinMCKnewFig1.png}
\caption{(a) The geometry of the spin-chain $M$-channel Kondo model (next-nearest neighbor coupling $J_2$ not shown). An impurity (left) and an image impurity (right) are both connected to the $M$ bulk channels but with different couplings $J_K$ and $J_K'$. The nearest-neighbor couplings in the bulk channel are always $J_1$. The next nearest-neighbor couplings $J_2$ are omitted in the figure for simplicity, see Eqs.~(\ref{eq:ham_bulk})-(\ref{eq:ham_image}) for details. The total number of spins is $M\times L + 2$.  The cut position is labeled by $x=1,\dots,L-1$. The entanglement entropy as a function of $x$ for $J_K=J_K'=J_1$, $J_2=J_2^c = 0.2412J_1$ obtained from DMRG simulation for (b) $M=3$ and (c) $M=4$. 
The uniform part $S_u$ is close to the prediction $M\times S^{\text{OBC}} + \ln g(M)$ 
where from Eq.~\eqref{eq:g_MCK} $g(3) = \varphi$ is the golden ratio and $g(4) = \sqrt{3}$.  
In (b) $L=100$ and DMRG maximum bond dimension  $\chi_{\text{max}}=4000$ while  
in (c) $L=40$ and $\chi_{\text{max}}=5000$.}
\label{fig1}
\end{figure}
The impurity entropy is defined as the difference between the (uniform parts of the) entanglement entropies with and without the impurity, where the entanglement cut at $x$ is sufficiently far from the boundaries~\cite{alkurtass2016spin2CK} [see \cref{fig1}(a)], 
\begin{equation}
\label{eq:Simp}
    S_{\text{imp}}(x) = S^{M\text{CK}}_u(x) - M\cdot S^{\text{OBC}}(x) .
\end{equation}
Here, $S^{\text{OBC}}(x)$ is the von Neumann entanglement entropy of a 1D Heisenberg chain subject to open boundary conditions which is given by the Cardy formula \cite{calabrese2004entanglement}:
\begin{equation}
\label{eq:cardy}
    S^{\text{OBC}}(x) = \frac{c}{6} \ln\qty[\frac{2L}{\pi} \sin\qty(\frac{\pi x}{L})] + \frac{s_1}{2} + \ln g_{\text{Cardy}}\,,
\end{equation}
where $c=1$ is the central charge, $s_1$ is a non-universal factor, and $g_{\text{Cardy}}=2^{-1/4}$ is the non-integer ground-state degeneracy associated with a Heisenberg chain \cite{affleck1992noninteger}.  Notably, for the two-channel model ($M=2$), when $J_K = J_1$, $S^{2\text{CK}}_u(x)$ is given by the Cardy formula for a periodic chain of length $2L+2$ with a cut position $2x + 1$ [see \cref{fig1}(a)],
\begin{equation} \label{eq:pbc_cardy}
     S^{\text{2CK}}_u(x) \!\! = \! S^{\text{PBC}}\!(x) \! = \!\frac{c}{3} \ln\qty{\!\frac{2L+2}{\pi} \sin\qty[\frac{\pi (2x+1)}{2L+2}]\!} + s_1.
\end{equation}
When $x\ll L$, the impurity entropy Eq.~(\ref{eq:Simp}) becomes \cite{alkurtass2016spin2CK}:
\begin{equation}
    S_{\text{imp}}(x) 
    \approx -2 \ln  g_{\text{Cardy}}  = \ln\sqrt{2}\,, \quad (M=2).
\end{equation}
The value of $g=\sqrt{2}$ coincides with the prediction of \cref{eq:g_MCK} for $M=2$ and matches the quantum dimension of Majorana fermions. That said, it is important to note that $g=\sqrt{2}$ is not unique to the single-impurity two-channel spin-chain Kondo model~\cite{tang2025xx2ck,tsvelik2013threeising,tsvelik2013magnetic}, much less to the two-impurity Kondo model \cite{Mitchell2CKin2IK2012,BayatScalingIQPT2017}. 

To go beyond $M=2$, one needs to first check if  generalized models with $M>2$ channels can give \cref{eq:g_MCK}.  Unfortunately, a uniform periodic limit employed in Eq.~\eqref{eq:pbc_cardy} is not known for $M > 2$ channels. 
Although multichannel spin chain Kondo physics has been studied in Refs.~\cite{oshikawa2006junction,CRAMPE2013526,
tsvelik2013threeising,buccheri2018circulator,buccheri2019chiral,giuliano2020tunable}, it is not clear how to generalize the models to arbitrary number of channels.  
In this paper, we aim to resolve this ambiguity by proposing a spin-chain geometry to realize Kondo physics for an arbitrary channel number $M$.  To accomplish this task, we introduce the notion of an ``image impurity''~\cite{ColemanIoffeTsvelik2CK1995} which reduces to the usual periodic boundary condition when $M = 2$ and the open boundary condition when $M=1$, see Fig.~\ref{fig1}(a). Using our geometry, we are able to realize multichannel Kondo physics using spin chains for the particularly interesting cases of $M=3,4$ channels [Fig.~\ref{fig1}(b-c)], obtaining respectively  ${g \approx 1.623}$ and ${g \approx 1.787}$, close to the predictions $g=\varphi \approx1.618$ and $g=\sqrt{3}\approx 1.732$ of \cref{eq:g_MCK}. 
We can attribute the differences to a finite-size error, see Fig.~\ref{fig2}(a). 
Finite-size scaling of total impurity spin also confirms that our model realizes MCK physics, see Fig.~\ref{fig2}(b).

\section{\blue{Numerical Impurity Entropy}}~
In this work, we use the entanglement entropy as a probe for the $M$-channel Kondo model. 
We use DMRG (TenPy~\cite{Tenpy,code}) to numerically find the ground state of Eq.~(\ref{eq:MCK_ham}) and extract the entanglement entropy $S(x)$. 
The finite system size $L$ induces oscillations in $S(x)$ which allow us to decompose $S(x)$ into a uniform and alternating term~\cite{laflorencie2006boundarycritical}, $  S(x) = S_u(x) + (-1)^x S_a(x)$. In this work, we focus on the uniform part $S_u(x)$ for our analysis of the entanglement structure~\cite{laflorencie2008kondo,alkurtass2016spin2CK}. Given the entropy profile $S(x)$, we can extract $S_u(x)$ as the average of two cubic fits for the upper blue dots (odd $x$) and lower blue dots (even $x$) \cite{sorensen2007impurityentanglement}, see Fig.~\ref{fig1}(b-c).
We then calculate the impurity entropy $S_{\text{imp}}(x)$ as in Eqs.~(\ref{eq:Simp})-(\ref{eq:cardy}). 
The case of $J_K=1$ approximately corresponds to the Kondo fixed point \cite{laflorencie2008kondo,alkurtass2016spin2CK} wherein the contribution to the impurity entropy is a consequence of the spin-chain geometry itself.

We first present our numerical values of $g$ for the $M=3$ and $M=4$ models.  We compute $g$ from $\exp\qty(S_{\text{imp}})$ at $x=L/2$ which corresponds to the closest agreement with \cref{eq:g_MCK} as shown in \cref{fig1}(b-c).  In computing $S_{\text{imp}}$, we also must compute $S^{\text{OBC}}$ which requires the determination of the non-universal constant $s_1$.  Fortunately, $s_1$ can be easily extracted by comparing DMRG results with \cref{eq:pbc_cardy} for a periodic XXX chain. Based on this $s_1$ value, we find that, at the Kondo fixed point ($J_K=1$), $g\approx 1.623$ for $M=3$ and $g\approx1.787$ for $M=4$.
Both these values are very close to the predictions of \cref{eq:g_MCK}, i.e., $g(M=3)=1.618$ and $g(M=4)=1.732$, even for the relatively small system sizes we used. The difference can be attributed to finite size effect as we will discuss below. The bond dimension $\chi$ needed is approximately $L^{\alpha\ c_{\text{eff}}}$ where $\alpha$ is a constant and $c_{\text{eff}}=M$ is the effective central charge. The computational cost is about $\mathcal{O}(L\times\chi^3)\sim \mathcal{O}(L^{3\alpha M +1})$~\cite{Tenpy}, which scales exponentially with $M$.

We seek to further support our proposed multichannel model by showing that it reproduces a similar scaling of $S_{\text{imp}}$ and $\xi_K$ (Kondo correlation length)~\cite{MitchellrealspaceRG2011} observed for the one- and two-channel models~\cite{laflorencie2008kondo, alkurtass2016spin2CK}.  
We performed an analysis of $S_{\text{imp}}$ as $J_K$ was varied for different values of the system size $L$ to extract $\xi_K$ by fitting to the scaling form $S_{\text{imp}}(x/L,x/\xi_K)$. 
The extracted $\xi_K$ values are then shown in the inset of Fig.~\ref{fig2}(a) where our empirically measured $\xi_K$ indeed appears to scale as $e^{a/J_K}$. The resulting scaling law for $S_{\text{imp}}$ is shown in Fig.~\ref{fig2}(a). 
In contrast to the Fermi-liquid power law $1/(x/\xi_K)$ observed in the spin-chain 1CK model \cite{sorensen2007impurityentanglement,sorensen2007kondoimpurity,laflorencie2008kondo} and the non-Fermi liquid power law $\ln(x/\xi_K)/(x/\xi_K)$ for the spin-chain 2CK model \cite{alkurtass2016spin2CK}, we find the non-Fermi liquid power law $1/(x/\xi_K)^{4/5}$ in our spin-chain 3CK model, which is more dominant than the Fermi liquid $1/(x/\xi_K)$ coming from each XXX chain. Both finite-size power laws match the finite-temperature correction to the multichannel impurity entropy \cite{ErikssonSimpcorrection2011,KondoSimpcorrectionCFT}, which is related to the specific heat \cite{kimura2021kondoabcd} when taking the temperature $T\to 1/(x/\xi_K)$. 
Fig.~\ref{fig2}(a) allows us to estimate the finite-size correction $\exp S_{\text{imp}}(x)-  g(M)$.   
For $M=3$, $g(3) = \varphi$ and the parameters of Fig.~\ref{fig1}(b), we can estimate this correction to be $\sim 0.01$ for $x=L/2$. Similar estimate for $M=4$ (albeit using the above $\xi_K$) gives $\sim 0.03\equiv \epsilon$, of the same order of magnitude as $S_{\text{imp}}-\ln(\sqrt{3})=\ln(\sqrt{3}+\epsilon)-\ln(\sqrt{3})\approx \epsilon/\sqrt{3}$ 
in Fig.~\ref{fig1}(c). Similar to the above finite-size scaling is found for the total impurity spin, discussed next.

\begin{figure}[tb]
\raggedright
\includegraphics[width=0.95\columnwidth]{PRLv2spinMCKnewFig2.png}
\caption{(a) The impurity entropy scaling plot of the spin-3CK model with $J_K'=1$ fixed, $J_K$ and total length $L\in[50, 60, 64, 70]$ varying. It shows non-Fermi liquid power law behavior (with power $4/5$) when approaching the predicted impurity entropy $\ln \varphi$ for two cut positions $x=L/5$ and $x=L/10$. 
(b) The scaling plot for the $z$-component of total spin of the impurity and the image impurity for spin-1CK (red), 2CK (blue) and 3CK (green) with $J_K'=J_K$. At 1CK, the power law is of Fermi liquid type $1/(L/\xi_K)^{2}$, but non-Fermi liquid for 2CK and 3CK.
Here $\xi_K\sim \exp(2.0/J_K)$. 
The Kondo couplings $J_K=J_K'\in[0.2, 1]$ and the $L$ values are $[30, 40, 50, 60, 70]$ for $M=1,2$ and $[10,20,30,40]$ for $M=3$.
} 
\label{fig2}
\end{figure}

\section{\blue{Total Impurity Spin}}~Another signature of MCK physics in our model is exhibited by the total spin of the impurity and image impurity. 
We first find that the total spin of the whole system is spin $0$ for all $0<J_K'=J_K\le1$ and $M=1,2,3$. 
This singlet formed by the impurity spin and the rest of the spins is similar to the electronic multichannel Kondo model where the impurity spin was connected to entanglement negativity~\cite{donghoon2021universalthermalentanglement}. The negativity was also numerically calculated for the spin-chain 1CK case~\cite{negativityspin1CK}. 
We calculate the total spin of the impurity and the image impurity, or equivalently the total spin of the $M\times L$ bulk spins because the whole system has spin 0. Due to the $\text{SU}(2)$ symmetry, it is sufficient to calculate the $z$-component,
\begin{equation} \label{eq:Sztot}
     \langle (\mathrm{S}^z_{\text{tot}})^2\rangle = \langle ( \mathrm{S}_0^z + \mathrm{S}_0^{'z})^2\rangle = \frac{1}{2} + 2 \langle \mathrm{S}_0^z \mathrm{S}_0^{'z} \rangle \,,
\end{equation}
where $\langle \mathrm{S}_0^z \mathrm{S}_0^{z} \rangle=\langle \mathrm{S}_0^{'z} \mathrm{S}_0^{'z} \rangle=1/4$ by $\text{SU}(2)$ symmetry.

We find that the total impurity spin Eq.~(\ref{eq:Sztot}) depends on the system size and Kondo coupling through the ratio  $L/\xi_K$. 
The finite-size scaling for 
$\langle (\mathrm{S}^z_{\text{tot}})^2 \rangle$  
is shown in Fig.~\ref{fig2}(b). 
In the $L/\xi_K \to \infty$ limit, $\langle (\mathrm{S}^z_{\text{tot}})^2\rangle \to 1/2$, which corresponds to uncorrelated impurity and  image impurity. 
In the bulk, we get exactly the same value for the $z$-component total spins, which suggests that two independent spin-1/2 Kondo clouds are formed in the bulk. Because $\xi_K$ is the Kondo-cloud size, we can think of it the following way: One of the Kondo clouds is for the impurity, the other is for the image-impurity and they are decoupled from each other because their sizes $\xi_K$ are very small compared to the chain length $L$.
In this limit, we find $\langle (\mathrm{S}^z_{\text{tot}})^2 \rangle$ to deviate from 1/2 in a Fermi liquid power law $1/(L/\xi_K)^2$ for $M=1$ but non-Fermi liquid power laws $1/(L/\xi_K)$ for $M=2$ and $1/(L/\xi_K)^{4/5}$ for $M=3$, see Fig.~\ref{fig2}(b). These finite-size corrections can be understood to arise in second-order perturbation theory due to the least irrelevant operator with scaling dimension $2$ for $M=1$ and $1+2/(M+2)$ for $M\ge 2$ \cite{affleck1995cftkondo,kimura2021kondoabcd}. 
A similar scaling behavior can be found in Ref.~\cite{donghoon2021universalthermalentanglement} for the negativity of the impurity and the bulk electrons. In the opposite limit $L/\xi_K \to 0$, 
the correlation between the impurity and the image-impurity mediated by their Kondo clouds becomes strong. Eventually, the impurity and the image-impurity form a singlet and so do $M\times L$ bulk spins.

\section{\blue{Effect of Anisotropy}}~Due to finite size effects, a next-nearest neighbor coupling $J_2$ may be needed to suppress a bulk marginal term \cite{laflorencie2008kondo}.  In general, $J_2^c$ will be a function of $\Delta$ [as will be $s_1$ in Eq.~\eqref{eq:cardy}], which we find numerically~\cite{SM}. Equipped with these values of $J_2(\Delta)$ and $s_1(\Delta)$, we can then compute the impurity entropy of our multichannel Kondo model, focusing on the uniform case $J_K = 1$.  
The cases $M=1,2$ correspond to open and periodic XXZ Heisenberg chains, where it is known that~\cite{affleck1998edgemagneticfield} $g_{\text{cardy}}(\Delta)=[2d(\Delta)]^{-1/4}$ with $d(\Delta)=1- \arccos(\Delta)/\pi$~\footnote{Here $d(\Delta)$ is related to the effective Luttinger liquid parameter $K=1/(2d)$~\cite{giamarchi2003quantum}}, and thus $g(M=2,\Delta) =\sqrt{2}[d(\Delta)]^{1/2}$. 
\begin{figure}[tb]
\centering
\includegraphics[width=0.98\columnwidth]{PRLv2spinMCKnewFig3v4.png}
\caption{(a) The effective ground state degeneracy $g=\exp(S_{\text{imp}})$ versus XXZ anisotropy parameter $d$ or $\Delta$ for $M=1,2,3,4$ and $L=50,50,50,30$ respectively. The power law $g\sim d^{(M-1)/2}$ is observed where the fit lines are for the $\Delta\leq 0$. The fit does not work well near $\Delta=1$, where it yields  $g \sim \sqrt{M}$ instead of Eq.~(\ref{eq:g_MCK}). (b) The decay exponent $p$ of the correlation $\langle S_0^z S_0^{\prime z}\rangle$ ($\sim 1/L^p$) and $\langle S_0^x S_0^{\prime x}\rangle$ (inset) versus $d$ or $\Delta$. The errors are given by the largest and smallest decay exponent fitted using two nearby even $L$ values from $92$ to $108$ for $M=1$, $42$ to $58$ for $M=2$ and $26$ to $42$ for $M=3$.
}
\label{fig3}
\end{figure}
We use DMRG simulations to find $g(M,\Delta)$ for $M=1,\dots,4$, shown in \cref{fig3}(a).  We find that $g(M,\Delta)\sim d(\Delta)^{\alpha} $ closely follows a power law with the exponent $\alpha \approx (M-1)/2$. The prefactor is almost independent of $\Delta$, with only weak dependence near isotropy $d = 1$, where Eq.~(\ref{eq:g_MCK}) is recovered.

The power-law behaviors of the $S_{\text{imp}}$ in Fig.~\ref{fig2}(a) are affected by the XXZ anisotropy~\cite{SM}. However, in order to better demonstrate the interplay of multichannel Kondo effects and the Luttinger liquid impurity physics~\cite{KaneFisher1992,GiulianoKondoweaklinkSpinChain2018,
TanEntanglementflowKaneFisher2025,KattelXXZ2025,RylandsQIluttinger2016,Sirker2008,Collura2013,ZhaoCriticalXXZdefects2006}, we show in Fig.~\ref{fig3}(b) that the decay exponent $p$ of $\langle S_0^z S_0^{\prime z}\rangle \sim 1/L^{p}$ [equivalent to $\langle S_0^x S_0^{\prime x}\rangle$ at $\Delta=1$ due to SU(2) symmetry, see the inset of Fig.~\ref{fig3}(b)] is almost a constant $p(\Delta)=2$ for $M=1$ and $p$ transitions from $p=2$ to $p=4/(M+2)$ as $d$ goes from $0$ to $1$ for both the $M=2,3$ cases. The $M=2$ case is exactly the periodic XXZ chain and the dominant decay of $\langle S^z S^{\prime z}\rangle$ is $1/L^2$ for $\Delta<0$ and $1/L^{1/d}$ for $\Delta>0$~\cite{gogolin2004bosonization}. Currently, there is no analytical analysis on the $M=3$ case but a similar transition also occurs near $\Delta=0$, see Fig.~\ref{fig3}(a)-(b). The decay exponent of $\langle S_0^x S_0^{\prime x}\rangle$ is closely $p(\Delta)=2 d(\Delta)$ for $M=1$ and $p(M,\Delta)=[4/(M+2)]d(\Delta)$ for both $M=2$~\cite{affleck1998edgemagneticfield} and $M=3$. As illustrated above, the exponent $4/(M+2)$ at $\Delta=1$ is determined by the scaling dimension of the least irrelevant operator $1+2/(M+2)$ in multichannel Kondo physics~\cite{affleck1995cftkondo,kimura2021kondoabcd}. With anisotropy $\Delta$, all the $M=1,2,3$ exponents are reduced by a factor $d(\Delta)$.

\section{\blue{Summary and Outlook}}~In summary, we have demonstrated for up to four channels that our proposed spin-chain multichannel Kondo model produces a boundary entropy $g$ consistent with \cref{eq:g_MCK}.
Thus, a fermionic background is not necessary for observing the multichannel Kondo effect. 
Therefore, the model can be experimentally realized in quantum computers and simulators~\cite{2023Natur.618..500K,Rosenberg_2024,PhysRevResearch.5.013183,PhysRevResearch.6.033107,trotzky2008time,simon2011quantum,PhysRevLett.115.215301,toskovic2016atomic,jepsen2020spin,PhysRevX.11.041025,sarma2025designbenchmarksemulatingkondo} where the scaling of impurity total spin, Eq.~(\ref{eq:Sztot}), might be a straightforward first step as it requires only two-qubit measurement. 
Note that our discussion is limited to $J_K \leq 1$ because the model with $J_K>1$ may enter non-Kondo phases~\cite{PhysRevB.109.174416}. 
 
Our numerical results pose open questions especially for the anisotropic $\Delta\neq1$ case. Generalizing to multiple channels the Kondo model in a one-channel Luttinger liquid~\cite{PhysRevLett.69.3378,FurusakiKondoLuttinger1994,KondoHelicalEdge2009,JukkaKondoHelicalEdge2016,GiulianoKondoweaklinkSpinChain2018,TanEntanglementflowKaneFisher2025} (possibly related to quantum Brownian motion~\cite{YiKane,Yi}) might explain the observations in Fig.~\ref{fig3}. Other future directions include exploring possible relation to two-impurity spin-chain Kondo models, extending the model to other quantum impurity models such as non-Hermitian spin-chain multichannel Kondo models~\cite{burke2025nonhermitiannumericalrenormalizationgroup}, and 
studying finite-temperature and dynamical (finite-frequency) response functions.

\section{\blue{Acknowledgments}}
We have benefited from discussions with A. Zhakenov, A. Tsvelik, C.-M. Jian, E. K\"onig, M. Ye, O. Starykh, P. Kattel, R. Wang, T. Sedrakyan, Y. Zhou and Y. Tang. 
Numerical calculations were performed by using the TeNPy Library~\cite{Tenpy}. 
This work was supported by the U.S. Department of Energy, Office of Science, National Quantum Information Science Research Centers, Quantum Science Center. 
J.G. would like to acknowledge the Office of Undergraduate Research at Purdue University for financial support.  He was also financially supported through the Joel Spira Summer Research Award.

\bibliography{apssamp}

\clearpage
\newpage

\setcounter{equation}{0}
\setcounter{figure}{0}
\setcounter{section}{0}
\setcounter{table}{0}
\setcounter{page}{1}
\makeatletter
\renewcommand{\theequation}{S\arabic{equation}}
\renewcommand{\thepage}{S-\arabic{page}}
\renewcommand{\thesection}{S\arabic{section}}
\renewcommand{\thefigure}{S\arabic{figure}}
\renewcommand{\thetable}{S-\Roman{table}}

\renewcommand\section{\@startsection{section}{1}{0pt}%
  {1.5ex plus 1ex minus .2ex}%
  {1ex plus .2ex}%
  {\centering\normalfont\bfseries}}
  

\begin{widetext}
\begin{center}
Supplementary materials on \\
\textbf{``Spin-chain multichannel Kondo model via image impurity boundary condition''}\\
Jordan Gaines$^{1}$, Guangjie Li$^{1,2}$, Jukka I. V{\"a}yrynen$^{1}$, \\ 
$^{1}$ \textit{Department of Physics and Astronomy, Purdue University, West Lafayette, Indiana 47907, USA}\\
$^{2}$ \textit{Department of Physics and Astronomy, University of Utah, Salt Lake City, Utah 84112, USA}
\end{center}

These supplementary materials contain details about $J_2(\Delta)$ and $s_1(\Delta)$ in Sec.~\ref{SMsec1} and the 3CK $g(\Delta)$ values for all $J_2=0$ in Sec.~\ref{SMsec2}. Finally, we investigate the influence of  anisotropy $\Delta$ on the impurity entropy scaling in Sec.~\ref{SMsec3},  and in Sec.~\ref{SMsec4} we evaluate spin-spin correlation functions between the impurity and bulk spins.

\section{The critical $J_2$ for $\Delta\neq 1$}\label{SMsec1}
To adequately study Kondo physics when $\Delta \ne 1$, we will need to determine empirical values for $J_2^c(\Delta)$.
By numerically comparing the entanglement entropy of the XXZ chain with different values of $J_2^c$ to \cref{eq:cardy}, we can find the $J_2^c$ value that most closely reproduces $g_{\text{Cardy}}$ and hence obtain a good estimate of $J_2^c$.  For each $J_2^c$, we also need to repeat our prescription for finding $s_1$ which will now also be a function of $\Delta$.  The resulting $J_2^c$ values are shown in Fig.~\ref{SMfig1} below. Note that these critical values $J_2^c(\Delta)$ are different from the dimer-fluid critical points \cite{nomura1993phase} and the Gaussian fixed points \cite{nomura1994critical} of the XXZ chain.
\begin{figure}[h]
\centering
\includegraphics[width=0.9\columnwidth]{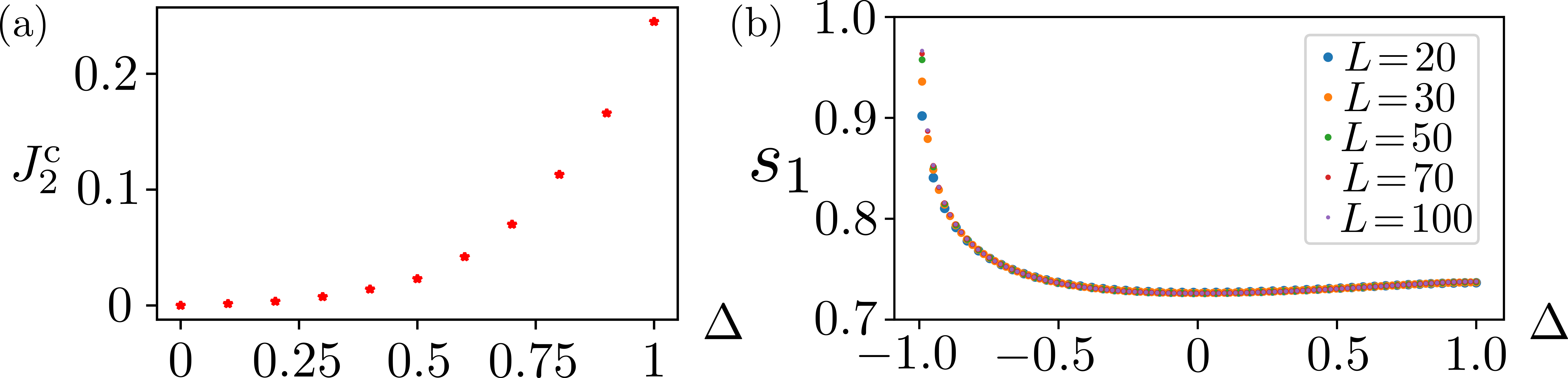}
\caption{(a) $J_2^c$ v.s. anisotropy $\Delta$ for $\Delta>0$. When $\Delta\leq0$, we use $J_2^c=0$. (b) $s_1$ v.s. anisotropy $\Delta$ for $L=20,30,50,70,100$ and $J_2=0$. }
\label{SMfig1}
\end{figure}

\section{The measured 3CK $g$ for all $J_2=0$}\label{SMsec2}
In this section, we compare the 3CK anisotropy results using $J_2=J_2^c(\Delta)$ with the same parameters except $J_2=0$ for all $\Delta$ and $L=20,50,70$. We observe that for the multichannel spin chain Kondo model the next-nearest neighbor coupling $J_2$ does not play a crucial role, see Fig.~\ref{SMfig2}.  
\begin{figure}[h]
\centering
\includegraphics[width=0.5\columnwidth]{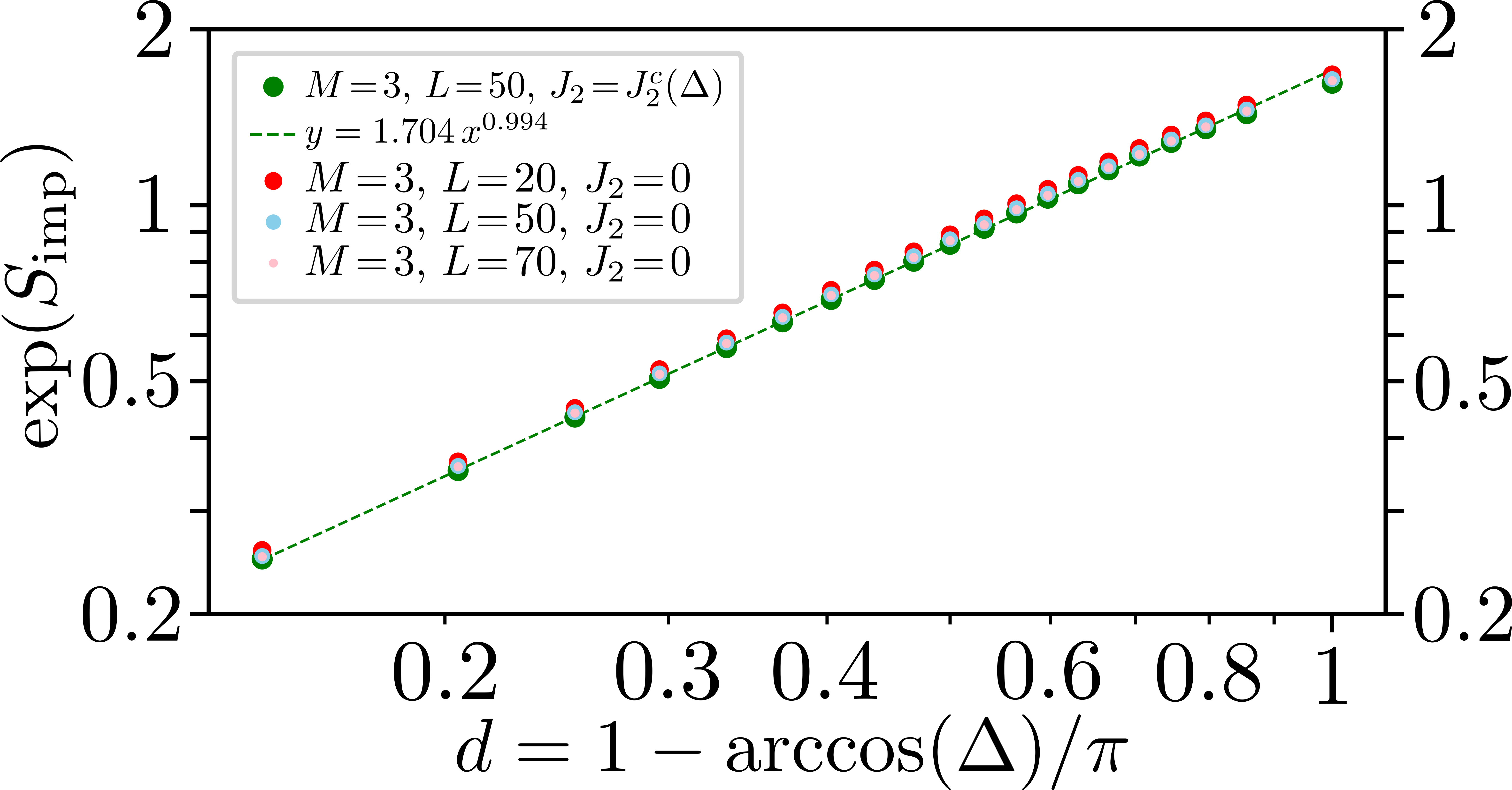}
\caption{The results with $J_2=J_2^c(\Delta)$ and $J_2=0$ are very close. The green points and dashed line have been shown in Fig.~\ref{fig3}.}
\label{SMfig2}
\end{figure}

\section{The $S_{\text{imp}}-\ln(g)$ power law v.s. anisotropy $\Delta$}\label{SMsec3}
We calculated the $S_{\text{imp}}(x)$ with $x=L/4$ by taking $\Delta\in[-0.9,-0.8,\dots, 1]$ and $L\in [44, 48, 52, 56]$ for $M=1,2$ and $L\in [28, 32, 36, 40]$ for $M=3$. By fitting the data we obtained, we get the exponent $p$ in $S_{\text{imp}}-\ln[g(M,\Delta)]\sim 1/L^p$ where $g(1,\Delta)=1$, $g(2,\Delta)=\sqrt{2}d^{1/2}$ and $g(3,\Delta)=\frac{\sqrt{5}+1}{2}d(\Delta)$ with $d(\Delta)=1-\arccos(\Delta)/\pi$. The $\Delta=1$ results for $M=1,3$ in the following Fig.~\ref{SMfig3} match the expected exponent $1$ for $M=1$ and $4/5$ for $M=3$. The $M=2,\Delta=1$ behavior is $\ln(L)/L$, not the power law $1/L$ and thus a lower effective power-law scaling $\sim0.6$ is observed. We did not find any simple dependence on $d$ and $\Delta$ for the exponent $p$, not like what we found in Fig.~\ref{fig3}(b) for the correlation functions.
\begin{figure}[h]
\centering
\includegraphics[width=0.5\columnwidth]{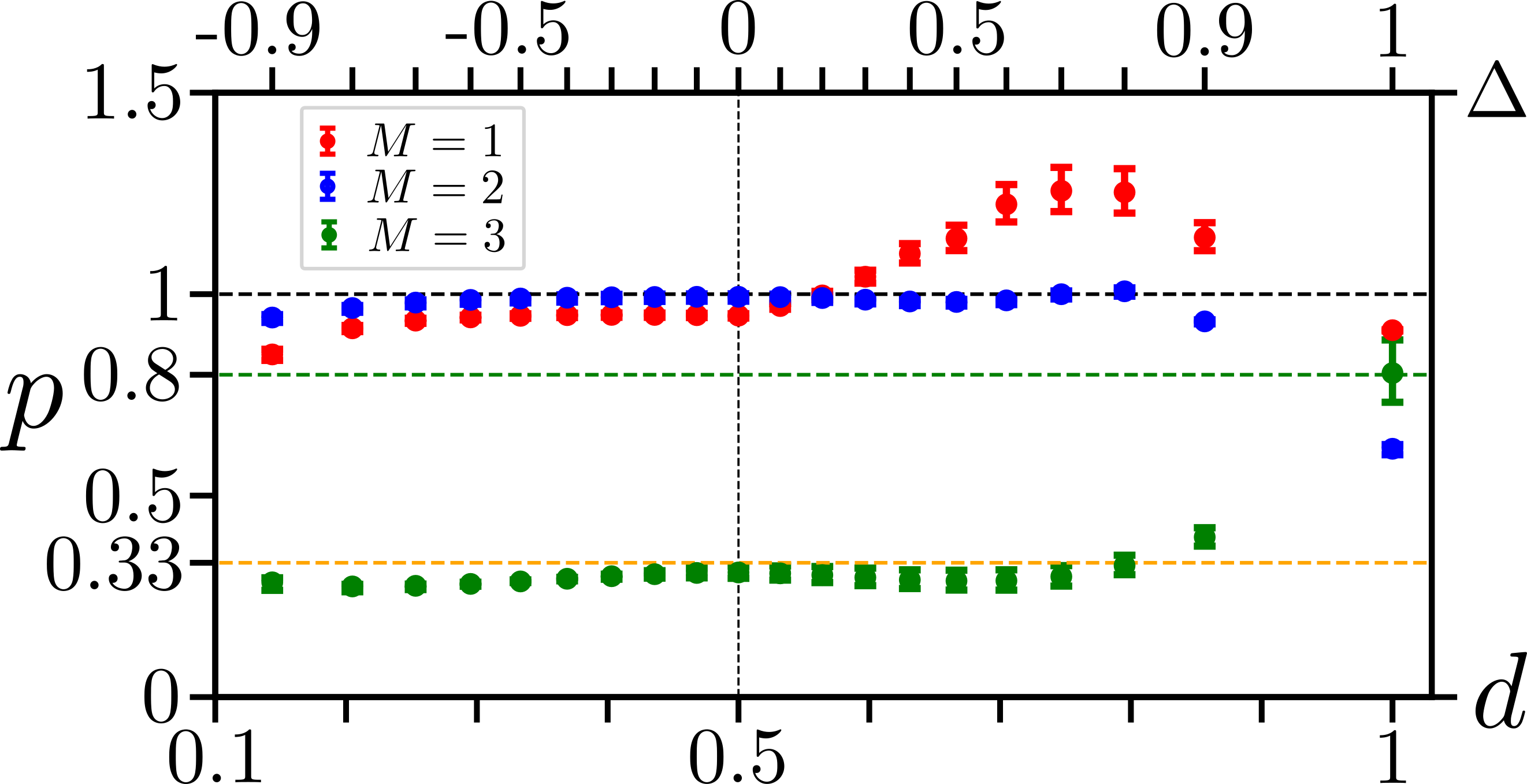}
\caption{The scaling exponent $p$ for $S_{\text{imp}}-\ln(g)\sim 1/L^p$ versus anisotropy parameter $d$ or $\Delta$ with $J_2=J_2^c(\Delta)$ for $M=1,2,3$.}
\label{SMfig3}
\end{figure}

\section{The $\langle S^x_0 S^{k=1,x}_r\rangle$ power law for different anisotropy parameters}\label{SMsec4}
In order to show the Luttinger liquid physics, we also calculate the correlation functions between the impurity spin $S_0$ and the spin in the $k$th channel and at position $r=1,\dots,L,L+1$ where $r=L+1$ denotes image-impurity. For example, the x-component can be written as $\langle S^x_0 S^{k=1,x}_r\rangle=\langle S^x_0 S^{k=1,x}_r\rangle_{U}+(-1)^r \langle S^x_0 S^{k=1,x}_r\rangle_{A}$ as a function of $r$ and the alternating part $\langle S^x_0 S^{k=1,x}_r\rangle_{A}$ is shown in Fig.~\ref{SMfig4}. The $M=2$ case (periodic XXZ chain) matches with the prediction in Ref.~\cite{affleck1998edgemagneticfield}.
\begin{figure}[h]
\centering
\includegraphics[width=0.92\columnwidth]{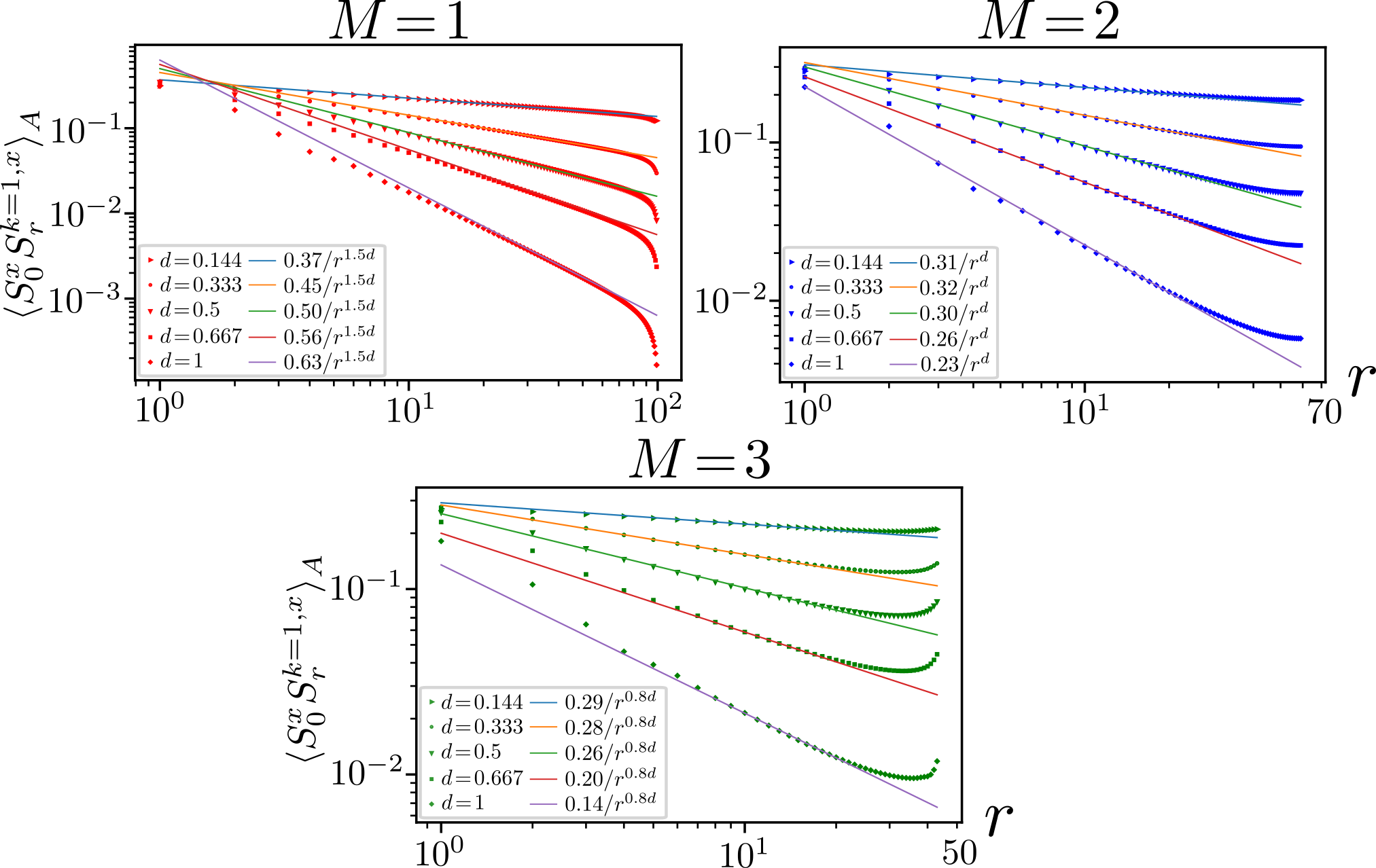}
\caption{A log-log plot of $\langle S^x_0 S^{k=1,x}_r\rangle_{A}$ v.s. $r$ for $M=1,2,3$ (red, blue, green) and $L=98,58,42$ respectively. The Luttinger-liquid scaling behavior is clearly seen from the power-law lines with exponents $(3/2)d, d, (4/5)d$ for $M=1,2,3$ and the Luttinger parameter is $K=1/(2d)$.} 
\label{SMfig4}
\end{figure}

\end{widetext}

\end{document}